\documentclass[12pt]{article}

\input{epsf.tex}

\newcommand{\E}{{\cal{E}}}

\newcommand{\be}{\begin{equation}}
\newcommand{\ee}{\end{equation}}
\newcommand{\bea}{\begin{eqnarray}}
\newcommand{\eea}{\end{eqnarray}}
\def\J#1#2#3#4{#1 {\it #2} {\bf #3} #4}
\def\PTP{\it Prog. Theor. Phys.}
\def\PRL{\it Phys. Rev. Lett.}

\def\PR{\it Phys. Rev.}

\def\APNY{\it Ann. Phys. (NY)}

\def\CQG{\it Class. Quantum Grav.}

\def\PLA{\it Phys. Lett. A}

\begin{document}
\title{Comment on ``The double--Kerr solution''\\
by W~B~Bonnor and B~R~Steadman}
\author{V~S~Manko\dag\, and
E~Ruiz\ddag}
\date{}
\maketitle

\vspace{-1cm}

\begin{center}
\dag Departamento de F\'\i sica,\\ Centro de Investigaci\'on y de
Estudios Avanzados del IPN,\\ A.P. 14-740, 07000 M\'exico D.F.,
Mexico\\ \ddag Area de F\'\i sica Te\'orica, Universidad de
Salamanca,\\ 37008 Salamanca, Spain
\medskip
\end{center}

\vspace{.1cm}

\begin{abstract}
\noindent Erroneous statements on the mathematical and physical
properties of the double--Kerr solution made in a recent paper
(2004 {\it Class. Quantum Grav.} {\bf 21}, 2723) are commented.
\end{abstract}

\medskip


\vspace{1.5cm}

\noindent The well--known Kramer--Neugebauer (KN) solution
\cite{KNe} for two Kerr black holes \cite{Kerr} has been analyzed
in a recent work \cite{BSt} by means of an approximation
procedure. Since the paper \cite{BSt} contains various erroneous
statements on the properties of the KN spacetime, some comments on
it seem appropriate.

\medskip

($i$) First of all, it is worthwhile pointing out that the BS
approximate potential (32) of \cite{BSt} lacks generality, that
invalidates the physical conclusions which Bonnor and Steadman
believe to have been drawn for the whole KN spacetime. This is
because their statement following formula (32) -- ``{\it As
expected, the solution has \underline{five} parameters: $m_s$,
$q_s$, $b$}'' -- is wrong. The general asymptotically flat
subfamily of double--Kerr spacetimes contains {\it six} arbitrary
real parameters, and the absence of one parameter in the BS scheme
is explained by the restrinctive character of the
reparametrization (12)--(15) employed in \cite{BSt} instead of the
original KN parametrization \cite{KNe}. Indeed, Kramer and
Neugebauer used in their paper {\it eight} real constants $K_l$,
$\omega_l$, $l=1,2,3,4$, of which the constants $K_l$ can be
subjected to the constraint $K_1+K_2+K_3+K_4={\rm const}$ due to
the liberty in the shift along the $z$--axis, while the four
constants $\omega_l$ give rise to the unphysical NUT parameter
which can be eliminated by an appropriate unitary transformation.
Therefore, in the general asymptotically flat case one is left
with {\it six} arbitrary real constants (8 parameters minus 2
constraints), and not {\it five} as in \cite{BSt}. A correct
reparametrization of the KN solution which does not lead to the
loss of one parameter was performed in \cite{DHo}.

Below we give a possible 6--parameter axis expression for the
Ernst potential \cite{Ernst} of the asymptotically flat
double--Kerr solution:
\be
\E(\rho=0,z)=\frac{z-b-M_1-i(A_1+\nu)}{z-b+M_1-i(A_1-\nu)}\cdot
\frac{z+b-M_2-i(A_2-\nu)}{z+b+M_2-i(A_2+\nu)}, \ee where the
parameters $M_i$ represent the masses of the Kerr particles, $A_i$
their angular momenta per unit mass, $b$ is the separation
constant, and $\nu$ can be associated with the angular momentum of
the part of the intermediate region.

\medskip

($ii$) It is important to clarify that Bonnor and Steadman expand
the KN solution not only in powers of masses, as they claim, but
also in powers of the angular momenta per unit mass since the
latter are implicitly defined in terms of the mass parameters.
This fact is a key point for understanding why the approximate
potential they obtained describes a specific system of black
holes, and not a system involving superextreme Kerr constituents.

\medskip

($iii$) It seems that the biggest confusion of the paper
\cite{BSt} is connected with the physical interpretation of the
region separating the particles in the double--Kerr solution.
Bonnor and Steadman take as granted that the black holes of the KN
solution are separated by a massless strut (see Fig.~1a), most
probably being unaware of the fact that the parameters they are
using have a rather formal character (their ``masses $m_1$,
$m_2$'' for instance are not really the masses of black holes) and
also describe the systems of not separated Kerr constituents,
e.g., as shown in Fig.~1b where a ``massive strut'' is present
(the two rods can also intersect partially). Moreover, one does
not know from the very beginning which situation describes the KN
solution, so that the correct choice of values of the parameters
is of paramount importance.

\begin{figure}[htb]
\centerline{\epsfysize=80mm\epsffile{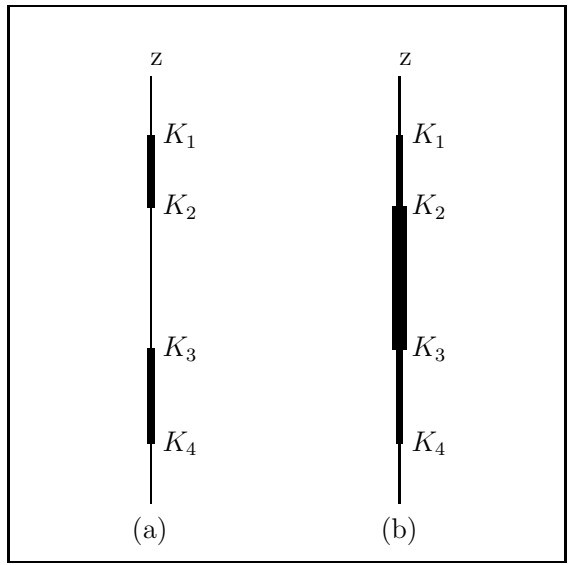}} \caption{Possible
configurations of two black holes in the KN solution.}
\end{figure}

Analyzing the case of two separated black holes, one should
remember that after having warranted the asymptotic flatness of
the KN solution, the next priority is to achieve
\be
\omega=0 \ee on the part $K_3<z<K_2$ of the $z$--axis, the
function $\omega$ entering the axisymmetric line element
\be
ds^2=f^{-1}[e^{2\gamma}(d\rho^2+dz^2)+\rho^2d\varphi^2]
-f(dt-\omega d\varphi)^2 \ee (the other two metric coefficients
are $f$ and $\gamma$). Fulfillment of the condition (2) converts
the region $\rho=0$, $K_3<z<K_2$ into a {\it massless line strut}
containing no closed timelike curves (that is why (2) is sometimes
called the axis condition \cite{DHo}).

The case $\omega\ne 0$ for $K_3<z<K_2$ describes the systems of
two overlapping Kerr subextreme constituents of the type shown in
Fig.~1b. In this case the intermediate region $\rho=0$,
$K_3<z<K_2$ possesses in general a non--zero mass and a non--zero
angular momentum.

In view of the above said, the physical interpretation given in
\cite{BSt} to the solution~I -- two Kerr particles supported by a
{\it massless} strut -- is erroneous because $\omega\ne 0$ between
the particles and hence the intermediate region in that solution
is not massless. Consequently, interpretation of the latter region
as ``{\it a massless spinning rod with angular momentum
$2m_1m_2q$}'' is erroneous too, and this can be easily illustrated
by calculating the Komar mass $M$ of any part of that rod with the
aid of the formula \cite{Tom}
\be
M=-\frac{1}{4}\omega[\psi(z=z_1)-\psi(z=z_2)], \quad z_1>z_2, \ee
where the functions $\psi$ and $\omega$ are defined by the
formulae (38), (41) and (45) of \cite{BSt} taken at $\rho=0$,
$|z|<b$, and $z_1$, $z_2$ are two arbitrary points on the rod.
Thus, for instance, choosing $m_1=m_2=1$, $b=3$, $q=0.6$ in the
latter formulae, one obtains from (4) that $M\approx -18.088$ for
$z_1=2.5$, $z_2=-2.5$.

Therefore, the BS solution~I belongs to the case of two
overlapping subextreme Kerr constituents (Fig.~1b) and its
intermediate region has some internal structure which gives rise
to the closed timelike curves due to the presence of the negative
mass. This solution is physically meaningless and is inapplicable
for the description of two separated particles. At the same time,
it cannot of course cast any doubt on the importance of the
general KN solution since one should restrict his consideration to
the physical range of the parameters and exclude unphysical
configurations.

There was a search for the equilibrium states of two Kerr
particles since 1980, and progress was achieved on the way of
extending the original KN solution \cite{KNe} to the case
involving superextreme Kerr constituents \cite{DHo}. At present
the double--Kerr balance problem can be considered as completely
solved \cite{MRu1,MRu2}, and the paper \cite{BSt} in no way
contributes either into the analysis of the genuine equilibrium
configurations of two spinning particles without a strut, or even
into shedding some new light on the systems composed of two Kerr
black holes.

\end{document}